\documentclass[pre,aps,superscriptaddress,twocolumn,floatfix]{revtex4-1}
\usepackage{dsfont}
\usepackage{mathtext}

\usepackage{mathtools}
\usepackage[usenames,dvipsnames]{xcolor}
\usepackage{amsmath}
\usepackage{amssymb,mathrsfs}
\usepackage{graphicx}
\usepackage{epstopdf}

\usepackage{mathtext}

\usepackage{bm}

\newcommand{\br}{{\bm r}}

\newcommand{\bs}{{\bm s}}

\newcommand{\cH}{\mathcal{H}} 
\newcommand{\PT}{\mathcal{PT}}

\newcommand{\tcH}{\tilde{\mathcal{H}}}

\newcommand{\RNum}[1]{\uppercase\expandafter{\romannumeral #1\relax}}

\begin{document}
	
	\title{Floquet Edge Multicolor Solitons}

	\author{Sergey~K.~Ivanov}
	\affiliation{Moscow Institute of Physics and Technology, Institutsky lane 9, Dolgoprudny, Moscow region, 141700, Russia}
	\affiliation{Institute of Spectroscopy, Russian Academy of Sciences, Troitsk, Moscow, 108840, Russia}
	
	\author{Yaroslav~V.~Kartashov}
	\affiliation{Institute of Spectroscopy, Russian Academy of Sciences, Troitsk, Moscow, 108840, Russia}
	\affiliation{ICFO-Institut de Ciencies Fotoniques, The Barcelona Institute of Science and Technology, 08860 Castelldefels (Barcelona), Spain}
	
	\author{Alexander~Szameit}
	\affiliation{Institute for Physics, University of Rostock, Albert-Einstein-Str. 23, 18059 Rostock, Germany}
	
	\author{Lluis~Torner}
	\affiliation{ICFO-Institut de Ciencies Fotoniques, The Barcelona Institute of Science and Technology, 08860 Castelldefels (Barcelona), Spain}
	
	\author{Vladimir~V.~Konotop}
	\affiliation{Departamento de F\'isica, Faculdade de Ci\^encias, Universidade de Lisboa, Campo Grande, Ed. C8, Lisboa 1749-016, Portugal}
	\affiliation{Centro de F\'isica Te\'orica e Computacional, Universidade de Lisboa, Campo Grande, Ed. C8, Lisboa 1749-016, Portugal}
	
	\begin{abstract}
		{Topological insulators are unique physical structures that are insulators in their bulk, but support currents at their edges which can be unidirectional and topologically protected from scattering on disorder and inhomogeneities. Photonic topological insulators can be crafted in materials that exhibit a strong nonlinear response, thus opening the door to the exploration of the interplay between nonlinearity and topological effects. Among the fascinating new phenomena arising from this interplay is the formation of topological edge solitons --- hybrid asymmetric states localized across and along the interface due to different physical mechanisms. Such solitons have so far been studied only in materials with Kerr-type, or cubic, nonlinearity. Here the first example of the topological edge soliton supported by parametric interactions in $\chi^{(2)}$ nonlinear media is presented. Such solitons exist in Floquet topological insulators realized in arrays of helical waveguides made of a phase-matchable $\chi^{(2)}$ material. Floquet edge solitons bifurcate from topological edge states in the spectrum of the fundamental frequency wave and remain localized over propagation distances drastically exceeding the helix period, while travelling along the edge of the structure. A theory of such states is developed. {It is shown that multicolor solitons in a Floquet system exists in the vicinity of (formally infinite) set of linear resonances determined by the {\em Floquet phase matching} conditions.} Away from resonance, soliton envelopes can be described by a period-averaged single nonlinear Schr\"odinger equation with an effective cubic nonlinear coefficient whose magnitude and sign depend on the overall phase-mismatch between the fundamental frequency and second harmonic waves. Such total phase-mismatch includes the intrinsic mismatch and the geometrically-induced mismatch introduced by the array, and its value reveals one of the genuine effects exhibited by the Floquet quadratic solitons. {Our results open fundamental new prospects for the exploration of a range of parametric frequency-mixing phenomena in photonic Floquet quadratic nonlinear media}.}
	\end{abstract}

	\maketitle
	
\section{Introduction}

The phenomenon of topological insulation, first introduced in solid-state physics, has grown into rapidly expanding interdisciplinary research concept, covering many areas of modern physics, where topological insulators and associated unique physical phenomena and applications have been demonstrated (see reviews~\cite{HasanKane-10,QiZhang-11}). For instance, topological insulators have been observed experimentally and analyzed theoretically in mechanical systems~\cite{SusstrunkHuber-15}, acoustics~\cite{HeGeSunChenLuLiuChen-16, PengQinZhaoShenXuBaoJiaZhu-16}, with cold atoms in optical lattices~\cite{ZhuDalibardDauphinGerbierLewensteinZollerSpielman-13, LiYeChenKartashovTornerKonotop-18}, atomic Bose-Einstein condensates~\cite{condensates01, GaliloLeeBarnett-17}, polaritons in microcavities ~\cite{polaritons01,polaritons02,polaritons03}, and in various photonic systems \cite{photonics01,photonics02,photonics03,photonics04}. The majority of experiments with topological systems, conducted so far, were performed in essentially linear regime, see, for instance, recent overviews on topological effects in photonics ~\cite{linreview01,linreview02}. At the same time, in some of the above mentioned physical systems, including photonic ones, the nonlinear effects arising upon increase of the amplitude of excitations may become strong enough to significantly alter topological phases or at least to notably affect propagation dynamics of the topological edge states.

Recent advances in the actively expanding field of nonlinear topological photonics in conservative and dissipative systems are summarized in~\cite{SmirnovaLeykamChongKivshar-20, OtaTakataOzawa-20, Rachel-18}. It was shown that self-action in topological insulators can lead to the formation of closed currents in the bulk of topological system \cite{LumerPlotnikRechtsmanSegev-13, MukherjeRechtsman-20}, that it stimulates modulational instabilities of the edge states~\cite{LeykamChong-16, LumerRechtsmanPlotnikSegev-16, KartashovSkryabin-16}, leads to rich bistability phenomena \cite{bistab01,bistab02}, and causes energy shifts for edge states and their hybridization with bulk modes \cite{DobrykhYulinSlobozhanyukPoddubnyKivshar-18}. Due to localization properties of the edge states nonlinear processes stemming from Kerr nonlinearity can be enhanced for them, leading to the efficient third-harmonic generation \cite{harmonic01,harmonic02}. Moreover, nonlinear effects may become sufficiently strong to induce topological phases in materials that are topologically trivial in the linear regime \cite{selfinduced01, selfinduced02, selfinduced03}, the phenomenon that only recently received experimental confirmation in a photonic system \cite{selfinduced04}.

One of the most striking manifestations of nonlinearity in topological insulators is the possibility of formation of topological edge solitons in them. They are hybrid objects localized near the edge of the insulator due to self-action, inheriting topological protection and moving along the edge of the insulator over considerable distances without spreading, in contrast to their linear counterparts. {Such topological edge solitons typically bifurcate from linear edge states provided that the dispersion of the topological system (for a given linear edge state) allows formation of localized excitations for a given type of nonlinear interaction. Since many topological materials and systems are intrinsically nonlinear, topological edge solitons are encountered in diverse areas of physics and therefore they represent a universal physical phenomenon. For instance, topological edge solitons have been introduced theoretically, and in some cases demonstrated experimentally in mechanical systems \cite{solmech01,solmech02}, in nonlinear topological electric circuits \cite{solelec01,solelec02}, in Bose-Einstein condensates with spin-orbit interactions \cite{solbec01}, and in topological systems governed by Dirac equation \cite{soldirac01}. In photonic systems, topological edge solitons have been shown to form in helical waveguide arrays \cite{LeykamChong-16, AblowitzCurtisMa-14, AblowitzCole-17, AblowitzCole-19, IvanovKonotopSzameitKartashov-20, IvanovKartashovSzameitMaczewskyKonotop-20, IvanovKartashovMaczewskySzameitKonotop-20}, and were very recently observed in anomalous topological insulator \cite{MukherjeeRechtsman-20}, in optically induced Su-Schrieffer-Heeger \cite{ssh01,ssh02} lattices, and studied in polaritonic systems \cite{KartashovSkryabin-16, GulevichYudinSkryabinIorshShelykh-17,LiYeChenKartashovFerrandoTornerSkryabin-18}. Edge solitons in optically induced lattices have been also observed in resonant atomic vapors \cite{vapor01} and topological edge solitons and frequency combs were also predicted to form in driven two-dimensional arrays of coupled ring resonators \cite{solcomb01}.}

%Topological edge solitons have been shown to form in helical waveguide arrays \cite{LeykamChong-16, AblowitzCurtisMa-14, AblowitzCole-17, AblowitzCole-19, IvanovKonotopSzameitKartashov-20, IvanovKartashovSzameitMaczewskyKonotop-20, IvanovKartashovMaczewskySzameitKonotop-20}, and were very recently detected experimentally in anomalous topological insulator \cite{MukherjeeRechtsman-20}, in optically induced Su-Schrieffer-Heeger \cite{ssh01,ssh02} lattices, as well as in polaritonic \cite{KartashovSkryabin-16, GulevichYudinSkryabinIorshShelykh-17, LiYeChenKartashovFerrandoTornerSkryabin-18} systems. Edge solitons in optically induced lattices have been observed in resonant atomic vapors \cite{vapor01}.

Nevertheless, in spite of the considerable current interest to the formation of self-sustained edge solitons in topological insulators, so far they have been studied only in systems with cubic or Kerr-type nonlinearities. Fundamentally important questions, whether such states can form due to parametric nonlinear interactions of several waves, e.g., in $\chi^{(2)}$ optical materials, and whether such solitons acquire in these materials new unexpected features, remain unaddressed. In this work, we, for the first time to our knowledge, predict that $\chi^{(2)}$ nonlinear topological insulators can support long-living edge solitons, develop their theory and, using multiple-scale approach, derive equations governing evolution of their envelopes and dictating their parameters. We show that the properties of such states are governed by the dispersion of the edge state in fundamental frequency wave, from which they bifurcate (due to specific character of $\chi^{(2)}$ system, where array impacts differently two frequency components, quasi-propagation constant spectra for two harmonics are different and may not overlap), and by the effective nonlinear cubic coefficient that resonantly varies with phase mismatch.

Due to the fact that one of the most powerful platforms for the realization of nonlinear topological phases in photonics is connected with Floquet systems \cite{MukherjeRechtsman-20, selfinduced04, MukherjeeRechtsman-20}, in the present work we consider Floquet topological insulator realized as an array of helical waveguides, inscribed in $\chi^{(2)}$ medium. Floquet topological insulators are unusual physical systems, where nontrivial topological phases appear due to periodic variations of system parameters in the evolution variable (time or propagation distance) \cite{Floquet01,Floquet02}. Theoretically proposed in semiconductor systems \cite{Floquet03}, Floquet insulators based on helical waveguide arrays have been successfully used for the illustration of various linear phenomena, including anomalous topological phases \cite{Floquet04,Floquet05,Floquet06}, topological currents in quasicrystals \cite{Floquet07}, topological Anderson insulators \cite{Floquet08,Floquet09}, and many others. Importantly, due to the dynamical ``time-periodic'' nature of the Floquet insulators, edge solitons in them always exhibit fast oscillations, following variations of insulator profile on each period, and can slowly radiate, but their envelopes can still be accurately described by the period-averaged equations that we derive here for $\chi^{(2)}$ materials. We show that in contrast to the conventional discrete and continuous solitons in straight $\chi^{(2)}$ waveguide arrays that are typically pinned to the array sites and are usually quickly trapped when set into motion across the structure \cite{chi2array01, chi2array02, chi2array03, chi2array04, chi2array05, chi2array06, chi2array07, chi2array08} {(with the exception in \cite{Susanto07}, where considerable mobility enhancement for such solitons was encountered)}, edge solitons in our system travel along its edge due to their topological nature, traversing hundreds of array periods with negligible radiative losses.

We show that one of the most representative new features of nonlinear Floquet topological systems is that they support a new type of phase matching mechanism, which we term {\em Floquet phase matching}, which must be satisfied for two essentially different band-gap structures corresponding to the fundamental frequency and second harmonic waves. Unlike in uniform $\chi^{(2)}$ systems, where resonant coupling conditions are achieved for the standard relation between the propagation constants of interacting waves, in Floquet systems resonant conditions occur in a Floquet band, and therefore they are met for a formally infinite set of propagation constants. Moreover, in topological Floquet systems the effective strength of the parametric interactions and, hence, the very existence of edge solitons crucially depends on the sign and magnitude of the effective nonlinear coefficient for the states involved. The effective nonlinear coefficient may change its sign upon variation of phase mismatch and array parameters, which allows to achieve the conditions for the formation of both bright and dark envelope solitons for the same linear edge state, corresponding to the same Bloch momentum, in sharp contrast to materials with Kerr nonlinearity. Our results open the possibility to study a rich set of parametric interactions in photonic Floquet quadratic nonlinear systems where the concept of Floquet phase-matching may be applicable. This includes other types of Floquet-mediated phase-matching schemes, parametric frequency-mixing processes, and topological and non-topological pumping schemes, to name a few possibilities. {We anticipate that our results are relevant to a number of systems, even beyond photonics, where parametric nonlinear interactions may be realized and play an important role in the evolution of excitations, including Bose-Einstein condensates in time-dependent external potentials creating topologically nontrivial phases, topological fiber-loop systems and arrays of driven coupled microresonators, as well as various nanophotonic systems.} 

\section{Theory of Floquet solitons in $\chi^{(2)}$ media}

\subsection{Model}

We address the propagation of light beams along the $z$-axis in a medium with a phase-matchable $\chi^{(2)}$ nonlinearity and an inhomogeneous refractive index distribution forming a honeycomb array of helical waveguides. In the paraxial approximation, the system is described by the coupled nonlinear equations for the dimensionless electric fields of the fundamental frequency (FF), $\psi_1$, and second harmonic (SH), $\psi_2$, waves \cite{chi2unif01,Belashenkov89,chi2unif02,chi2array09}:
\begin{align}  
\label{mainFF}
i\frac{\partial \psi_1}{\partial z}&=-\frac{1}{2}\nabla^2 \psi_1-V(\br,z)\psi_1-\psi_1^*\psi_2\;,
\\
\label{mainSH}
i\frac{\partial \psi_2}{\partial z}&=-\frac{1}{4}\nabla^2 \psi_2+\beta\psi_2-2V(\br,z)\psi_2-\psi_1^2.
\end{align}
Here $\nabla=(\partial_x,\partial_y)$, ${\bm r}=(x,y)$ is normalized to the characteristic transverse scale $a$; the propagation distance $z$ is scaled to the diffraction length $\kappa_1 a^2$; $\psi_1=[2\pi\omega_0^2\chi^{(2)}a^2/c^2]E_1$ and $\psi_2e^{i\beta z}=[2\pi\omega_0^2\chi^{(2)}a^2/c^2]E_2$ are the dimensionless complex amplitudes of the FF (at the frequency $\omega_0$) and SH (at the frequency $2\omega_0$) waves, $E_{1,2}$ are the dimensional electric field amplitudes of the FF and SH waves; $\chi^{(2)}$ is the relevant second-order susceptibility for type \RNum{1} phase-matching, whose value depends on the particular crystal and its spatial orientation \cite{chi2gen1,chi2gen2}, $\kappa_1=n_1(\omega_0)\omega_0/c$ and $\kappa_2=n_2(2\omega_0)2\omega_0/c$ are the wavenumbers of the FF and SH waves at frequencies $\omega_0$ and $2\omega_0$; $\beta=(2\kappa_1-\kappa_2)\kappa_1 a^2$ is the normalized phase mismatch.

Due to the difference of carrier frequencies of FF and SH waves, the optical potential $V(\br,z)$ describing the array is approximately two times stronger in the equation for SH wave field \cite{chi2array06,chi2array07,chi2array09}. This potential is $L$-periodic along the $y$-direction and $Z$-periodic along the $z$-direction (here $z\geq0)$: $V(\br+L\textbf{j},z) = V(\br,z+Z) = V(\br,z)$. The array is composed from identical helical waveguides [see schematic illustration in Fig.~\ref{figure1}(a)] of width $\sigma$, placed in the nodes $\br_{nm}$ of the honeycomb grid $V(\br,z)=p\sum_{mn} \exp{(-[\br-\br_{nm}-\bs(z)]^2/\sigma^2)}$, where $p=\max(\delta\chi^{(1)})\,2\pi(\omega_0 a/c)^2 $ is the depth of the waveguide array created by the modulation of linear dielectric susceptibility $\delta\chi^{(1)}$, and $\bs(z)=\rho\left(\sin(\omega z),\,\cos(\omega z)-1\right)$ describes helical trajectory of each waveguide with the Floquet period $Z=2\pi/\omega$ and radius $\rho$. The $y$-period of such array is $L=\sqrt{3}d$, where $d$ is the separation between neighbouring waveguides. Since we are interested in solitons appearing at the edge of this topologically nontrivial Floquet structure, the array is truncated along the $x$-axis to form two zigzag edges.

It should be stressed that such truncated arrays of helical waveguides can be fabricated using well-established laser writing technology in transparent $\chi^{(2)}$ materials, such as $\textrm{LiNbO}_3$ \cite{chi2arraywriting1,chi2arraywriting2,chi2waveguides}. 

\subsection{Spectrum of the linear array}

\begin{figure*}
	\centering
	\includegraphics[width=\linewidth]{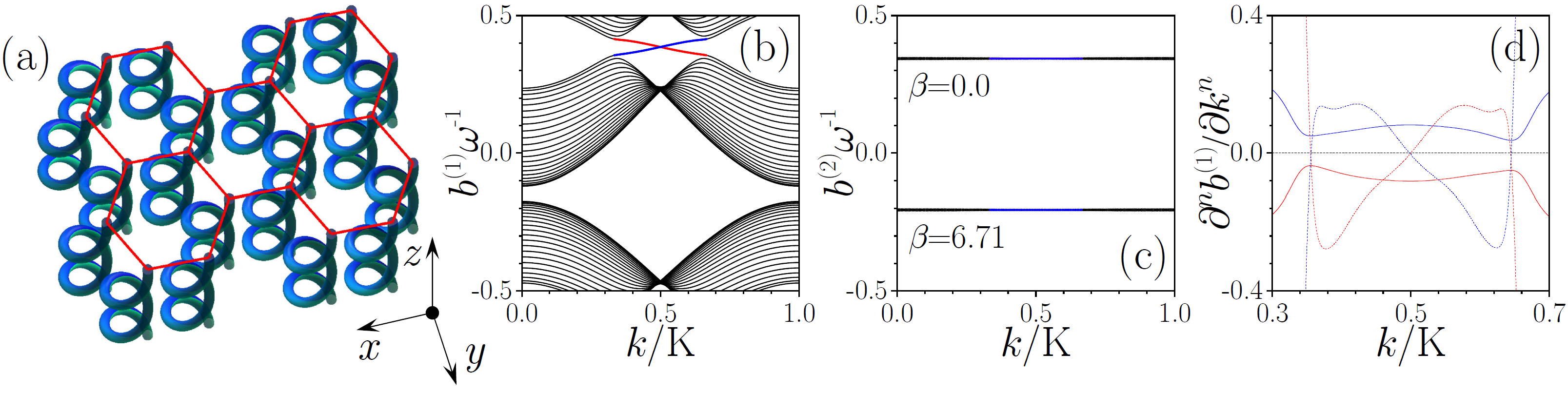}
	\caption{(a) Schematic representation of helical waveguide array inscribed in $\chi^{(2)}$ nonlinear medium. Quasi-propagation constants for FF wave $b^{(1)}$ (b) and SH wave $b^{(2)}$ (c) (in this case we superimpose band structures at $\beta=0$ and $\beta=6.71$ on the same plot because they are identical except for the vertical shift) versus normalized Bloch momentum $k/\textrm{K}$. (d) First-order derivatives $\partial b^{(1)}/\partial k$ (solid lines) and second-order derivatives $\partial^2 b^{(1)}/\partial k^2$ (dashed lines) for the edge states in the FF wave spectrum. Color coding correspond to panel (b). In all cases $p=11$, $d=1.7$, $\sigma=0.4$, $\rho=0.5$, $Z=8$.}
	\label{figure1}
\end{figure*}

First, we address the spectrum of modes supported by a linear waveguide array and described by the system (\ref{mainFF}), (\ref{mainSH}) with nonlinear terms set to zero. Due to the Floquet and Bloch theorems, the eigenmodes propagating in the $z$-direction can be searched as ($j=1,2$)~\cite{IvanovKonotopSzameitKartashov-20}:
\begin{align}
\psi_j(\br,z) = \phi^{(j)}_{\nu k}(\br,z)e^{ib^{(j)}_{\nu k}z} = u^{(j)}_{\nu k}(\br,z)e^{ib^{(j)}_{\nu k}z+iky}   
\end{align}
where $b_{\nu k}^{(1,2)} \in [-\omega/2,+\omega/2)$ are the quasi-propagation constants, $\nu$ is the band index, $k\in[-\textrm{K}/2,+\textrm{K}/2)$ is the Bloch momentum, with $\textrm{K}=2\pi/L$ being the width of the Brillouin zone,
\begin{align}
\phi^{(j)}_{\nu k}(\br,z)&=\phi^{(j)}_{\nu k}(\br,z+Z)
\\
u^{(j)}_{\nu k}(\br,z)&=u^{(j)}_{\nu k}(\br+L\hat{\bf j},z)=u^{(j)}_{\nu k}(\br,z+Z)
\end{align}
and the functions $\phi^{(j)}_{\nu k}(\br,z)$ are found from the two (uncoupled) linear eigenvalue problems
\begin{align}
\label{mainLin}
%\label{eqFF}
i\frac{\partial \phi_{\nu k}^{(j)}}{\partial z}-\cH_j\phi_{\nu k}^{(j)}=b_{\nu k}^{(j)}\phi_{\nu k}^{(j)}, \qquad j=1,2
%-\frac{1}{2}\nabla^2 \psi_1-V(\br,z)\psi_1\;,
%\\
%\label{eqSH}
%i\frac{\partial \phi^{(2)}}{\partial z}&=\cH_2\phi^{(1)}
%-\frac{1}{4}\nabla^2 \psi_2+\beta\psi_2-2V(\br,z)\psi_2.
\end{align}
where 
\begin{align}
\label{H12}
\cH_1=-\frac{1}{2}\nabla^2-V(\br,z), \quad 
\cH_2=-\frac{1}{4}\nabla^2+\beta-2V(\br,z). 
\end{align}

For subsequent consideration we define the inner product  
\begin{equation}
\label{inner}
(f(\cdot,z),g(\cdot,z))=\int_{S}f^*(\br,z) g(\br,z)d\br 
\end{equation}
where $S$ is the area of the array (notice that the inner product is defined for two functions taken at the same distance $z$). We also use the fact that states with different Bloch momenta $k$ are mutually orthogonal, while states from different bands or topological branches, say $\nu$ and $\nu'$, with equal Bloch momenta $k=k'$ have the following property
\begin{align}
(\phi_{\nu' k}^{(j)}(\cdot,z),\phi_{\nu k}^{(j)}(\cdot,z))=(u_{\nu' k}^{(j)}(\cdot,z),u_{\nu k}^{(j)}(\cdot,z))=\delta_{\nu\nu'}.
%=\frac{2\ell_y}{L}\left(u_{\nu' k},u_{\nu k}\right)_L
\end{align}  
valid for states considered at the same propagation distance $z$.

To obtain the linear spectra, i.e., the dependencies of the quasi-propagation constants $b_{\nu k}^{(j)}$ on Bloch momentum $k$, we use a propagation-projection method in which one first obtains the $z$-independent profiles of Bloch modes $\phi^{(j)}_{\nu k}$ from the two upper bands of the honeycomb array with straight waveguides (i.e., with $\rho=0$) using a plane-wave expansion method, and then propagates each of them exactly on one period $Z$ of the array with helical waveguides $(\rho\neq 0)$. By projecting, the resulting output fields $\psi^{(j)}_{\nu k}(z=Z)$ on all input modes $\phi^{(j)}_{\nu k}(z=0)$ one obtains a monodromy matrix, whose eigenvalues in the form of characteristic multipliers $e^{ib_{\nu k}^{(j)} Z}$ allow to extract the quasi-propagation constants $b_{\nu k}^{(j)}$ for the helical structure. Bloch modes of the helical array are then constructed using linear combinations $\sum_{\nu^\prime} c^{(j)}_{\nu \nu';k}\phi^{(j)}_{\nu' k}$, where $c^{(j)}_{\nu \nu';k }$ are proper elements of the matrix containing eigenvectors of the monodromy matrix. 

Examples of the Floquet spectra of truncated helical array obtained in this way for the FF [$j=1$, Fig.~\ref{figure1}(b)] and SH [$j=2$, Fig.~\ref{figure1}(c)] waves are presented in  Fig.~\ref{figure1}. Further, for simplicity in all plots we omit subscripts in $b_{\nu k}^{(j)}$. Waveguide rotation breaks the $z$-inversion symmetry and leads to opening of a topological gap in the spectrum of the FF wave (its width increases with $\rho$ up to certain critical value of the rotation radius), where two branches of the edge states are seen between the Dirac points $\textrm{K}/3 < k < 2\textrm{K}/3$ [Fig.~\ref{figure1}(b)]. Red branch corresponds to the states localized on the left edge and characterized by current in the positive $y$-direction (further we will use namely this branch for soliton construction), while blue branch corresponds to the states from the right edge that are associated with current in the negative $y$-direction. The black curves in the spectrum are associated with delocalized bulk modes. Because potential depth in Eq. (\ref{mainSH}) for the SH wave is nearly two times larger than that for the FF wave, while diffraction coefficient is two times smaller, the two bands with very narrow topological gap between them in the SH wave spectrum are nearly flat and appear practically as a line in Fig.~\ref{figure1}(c). Thus, almost the entire longitudinal Brillouin zone for SH wave $b^{(2)} \in [-\omega/2,+\omega/2)$ represents a forbidden gap -- a peculiarity of the spectrum especially relevant for edge soliton formation explored below. Notice that increasing phase mismatch leads to simple shift of the above mentioned nearly flat bands along the $b^{(2)}$ axis -- see spectra for $\beta=0$ and $\beta=6.71$ superimposed on the same Fig.~\ref{figure1}(c). Examples of the dependencies of the derivatives $\partial b^{(1)}/\partial k$ (characterizing group velocity) and $\partial^2 b^{(1)}/\partial k^2$ (characterizing dispersion) for the edge states in FF wave spectrum are illustrated in Fig.~\ref{figure1}(d). Remarkably, dispersion $\partial^2 b^{(1)}/\partial k^2$ can have opposite signs, depending on the $k$ value.

\subsection{Nonlinear model}

Turning now to the nonlinear case with parametric interactions included in Eqs. (\ref{mainFF}), (\ref{mainSH}) we recall that in a quadratic medium three different scenarios of bifurcation of the nonlinear families from linear ones may be encountered~\cite{MoreiraAbdullaevKonotop-12,MorAbKoYu2012,MorKonMal2013}: (i) FF and SH waves are of the same order (this case is realized, for example, when edges of the allowed bands in spectra of both fields coincide); (ii) SH field remains finite in the linear limit (this case requires the presence of the confining potential for SH wave); and (iii) in the linear limit FF and SH fields scale as $\psi_2=\mathcal{O}(\psi_1^2)$ (this is typical for the cascading limit in uniform medium or for periodic medium, when the propagation constant of the phase-matched SH wave falls into forbidden gap, while FF wave can propagate freely). As it follows from the comparison of spectra in panels (b) and (c) of Fig.~\ref{figure1}, namely the third scenario is characteristic for our system. In this case, in the linear limit the FF wave profile is described by Eq. (\ref{mainLin}) for $j=1$, while SH wave profile is described by Eq.~(\ref{mainSH}), which is a linear equation for $\psi_2$ with fixed $\psi_1^2$ term. 
This regime resembles the cascading limit, however it must be properly appreciated that the Floquet nature of the system imposes that the SH wave is driven by the periodically varying with $z$ FF wave, while the array introduces a geometrically-induced phase-mismatch between both waves. Therefore, the global phase-matching resonance between the waves will be shifted in the Floquet system relative to the homogeneous case.

We are interested in the family of Floquet edge solitons bifurcating under the action of nonlinearity from a linear topological edge state in the FF wave spectrum. We suppose that this state has band index $\alpha$ and corresponds to certain Bloch momentum $k$. To describe such solitons we employ the multiple-scale expansion~\cite{IvanovKonotopSzameitKartashov-20}. To this end, we introduce a formal small parameter $0<\mu\ll 1$ as well as scaled variables $y_j=\mu^jy$ and $z_j=\mu^j z$. These variables are considered independent and thus allow for representations of operators (\ref{H12}) in the form
\begin{align}
\label{Hamil1}
\cH_1=\tcH_1^{(0)}+\mu\tcH_1^{(1)}+\mu^2\tcH_1^{(2)}+\cdots
\end{align}
\begin{align}
\label{Hamil2}
\cH_2=\tcH_2^{(0)}+\cdots
\end{align}
where $\tcH_{1,2}^{(0)}$ are identical to $\tcH_{1,2}$, defined in~(\ref{H12}), with $\br$ replaced by $\br_0$, $\tcH_1^{(1)}=-\partial_{y_0}\partial_{y_1}$ and $\tcH_1^{(2)}=-\partial_{y_0}\partial_{y_2}-(1/2)\partial_{y_1}^2$. When searching for the profile of the Floquet edge soliton, we represent its FF component using the expansion
\begin{align}
\label{expan1}
\psi_1&=\mu e^{ibz_0} \left[A(y_1,z_1)\phi_{1}+\mu\rho_1+\mu^2\rho_2+\mathcal{O}(\mu^3)\right]
\end{align}
while its SH component is presented in the form
\begin{align}
\label{expan2}
\psi_2=\mu^2 e^{2ibz_0}A^2(y_1,z_1)\phi_{2}+\mathcal{O}(\mu^3),
%\left[A_1(y_1,z_1)\phi_{0}+\mu\phi^{(1)}+\mu^2\phi^{(2)}\right]
\end{align}
where slowly varying soliton amplitude $A(y_1,z_1)$ is introduced that defines the evolution of both fields. Notice that this expansion agrees with the scaling $\psi_2=\mathcal{O}(\psi_1^2)$ discussed above. Hereafter, in the arguments of a function we show only the most rapid variables and assume that the function can depend also on all respective slow variables (i.e., $A(y_1,z_1)$ may depend also on $y_2$ and $z_2$, but not on $y_0$ or $z_0$). To shorten the notations, we denoted $b_{\alpha k}^{(1)}=b$ and suppressed the double subscript for the carrier edge state $\phi_{\alpha k}^{(1)}\equiv\phi_{1}$ in the expression for FF profile. Notice that the function $\phi_{2}$ describing carrier state in the SH wave is not a SH eigenmode of helical array and its profile will be found later. The functions $\rho_{1,2}$ depend on all variables and they also will be found in the course of expansion. Our goal is to derive the equation describing evolution of the slowly varying amplitude $A(y_1,z_1)$.

Substituting this expansion into Eqs. (\ref{mainFF}) and (\ref{mainSH}) and keeping the terms up to the $\mu^2$-order in (\ref{mainFF}) and only the leading-order terms in (\ref{mainSH}) one obtains
\begin{align}
\label{exp-eq1}
i \partial_{z_0} (\rho_1&+\mu\rho_2)-b(\rho_1+\mu\rho_2) 
\nonumber \\
&\quad+i \partial_{z_1} \left(A\phi_{1}+\mu\rho_1\right)  +i \mu \phi_{1} \partial_{z_2}A
\nonumber \\
&=\tcH_1^{(0)}
\left(\rho_1+\mu\rho_2\right)
+\tcH_1^{(1)}
\left(A\phi_{1}+\mu\rho_1\right)
\nonumber \\
&\quad+\mu\tcH_1^{(2)}A\phi_{1}
+\mu\phi_{1}^*\phi_{2} |A|^2A ,
\\
\label{exp-eq2}
-i\partial_{z_0}\phi_2&+2b\phi_2 +\cH_2^{(0)}\phi_2=\phi_1^2.
\end{align}
The functions $\rho_{j}$ can be searched in the form of the expansions over the Floquet-Bloch states $\phi_{\nu k}^{(1)}$. For $j=1$ we have 
\begin{eqnarray}
\label{expan}
\rho_1=\sum_\nu B_{\nu k}(y_1,z_0)\phi_{\nu k}^{(1)} 
\end{eqnarray}
where the sum is over the band index only~\cite{IvanovKonotopSzameitKartashov-20}, and the coefficients $B_{\nu k}(y_1,z_0)$ are to be found. In the leading order, Eq.~(\ref{exp-eq1}) is reduced to
\begin{equation}
\label{exp-eq1-1}
i \partial_{z_0} \rho_1-b\rho_1 
+i(\partial_{z_1} A)\phi_{1} 
%\nonumber \\
=\tcH_1^{(0)}
\rho_1
-(\partial_{y_1}A)\partial_{y_0}\phi_{1}  
\end{equation}
Substituting the expansion (\ref{expan}) into this equation one obtains
\begin{align}
\label{first-order}
i\phi_1 \partial_{z_1}A  +(\partial_{y_1}A)\partial_{ y_0}\phi_1
%\nonumber \\
= \sum_\nu\phi_{\nu k}^{(1)}\left[-i\partial_{z_0}   +  (b-b_{\nu k})\right]B_{\nu k} 
\end{align}
Since the consideration at this order is similar to the one described in~\cite{IvanovKonotopSzameitKartashov-20}, we omit the details passing directly to the results. To obtain the dependence of the envelope on the slow variables $y_1$ and $z_1$ we project (\ref{first-order}) on $\phi_1$ and perform $z$-averaging over one helix period $Z$ (defined as $\langle f\rangle_Z=Z^{-1}\int_{0}^{Z}f(\br,z)dz$). This gives $A\equiv A(Y; z_2,x_2)$, where $Y=y_1-v z_1$ and the group velocity $v=-b^\prime$ is dictated by the group velocity of corresponding carrier FF edge state with index $\alpha$ at momentum $k$ (hereafter a prime stands for the derivative with respect to $k$, i.e., $v=-\partial b_{\alpha k}^{(1)}/\partial k$). $Z$-periodic expansion coefficients $B_{\alpha k}$ are given by
\begin{eqnarray}
\label{Ax1}
B_{\alpha k} 
=(\partial_{y_1}A)\int_{0}^{z}\left[\langle h\rangle_T-h(\zeta)\right]   d\zeta,
\end{eqnarray}
where we defined a $Z$-periodic function
\begin{align}
\label{period-h}
h(z+Z)=h(z):= -i\left(\phi_{1},\partial_{y}\phi_{1}\right).  
\end{align}
Other expansion coefficients $B_{\nu k} $ are obtained by projecting (\ref{first-order}) over $\phi_{\nu k}^{(1)}$ with $\nu\neq \alpha$ and performing $z$-averaging. Proceeding to the next order of $\mu$ in (\ref{exp-eq1}) and following main steps described in~\cite{IvanovKonotopSzameitKartashov-20} we arrive to the nonlinear Schr\"{o}diger (NLS) equation for the slowly varying amplitude:
\begin{eqnarray}
\label{NLS}
i\frac{\partial A}{\partial z}=\frac{b^{\prime\prime}}{2}
\frac{\partial^2A}{\partial Y^2}-\chi |A|^2A,
\end{eqnarray}
where
\begin{eqnarray}
\label{chi}
\chi=\langle ( \phi_1^2,\phi_2)\rangle_Z
\end{eqnarray}
is the effective cubic nonlinear coefficient and we returned to the physical variables by setting $\mu=1$ (i.e., in (\ref{NLS}) $Y=y-vz$). Thus, in the regime considered here, the slowly varying amplitude $A$ of the FF component of the Floquet edge soliton obeys the nonlinear Schr\"odinger equation. The  dispersion in this equation is defined by the dispersion of the linear FF edge state from which the FF component bifurcates. The effective cubic nonlinear coefficient depends both on the FF edge state $\phi_1$ and on the SH $\phi_2$ generated by the FF wave. 

%It is instructive to compare the derived NLS equation (\ref{NLS}) with that emerging in the cascading limit associated with large phase mismatch $\beta$ in a homogeneous medium. To this end we multiply (\ref{exp-eq2}) by $(\phi_1^*)^2$ and apply averaging over one $Z$-period (all variables are fast here and the subscript $0$ is omitted). After some algebra, one obtains
%\begin{align}
% \label{transformation}
%\langle ( \phi_1^2,\phi_1^2)\rangle_Z 
%=-\beta\chi-\frac{3}{2}\langle(\phi_1\nabla^2\phi_1,\phi_2)\rangle_Z-\frac{1}{2}\langle((\nabla %\phi_1)^2,\phi_2)\rangle_Z
%\end{align}
%One can see that at large enough $\beta$ one obtains the expression for the conventional cascading limit $-\beta\chi\approx \langle ( \phi_1^2,\phi_1^2)\rangle_Z$.

\subsection{Approximate effective cubic nonlinearity}

\begin{figure}
	\centering
	\includegraphics[width=\linewidth]{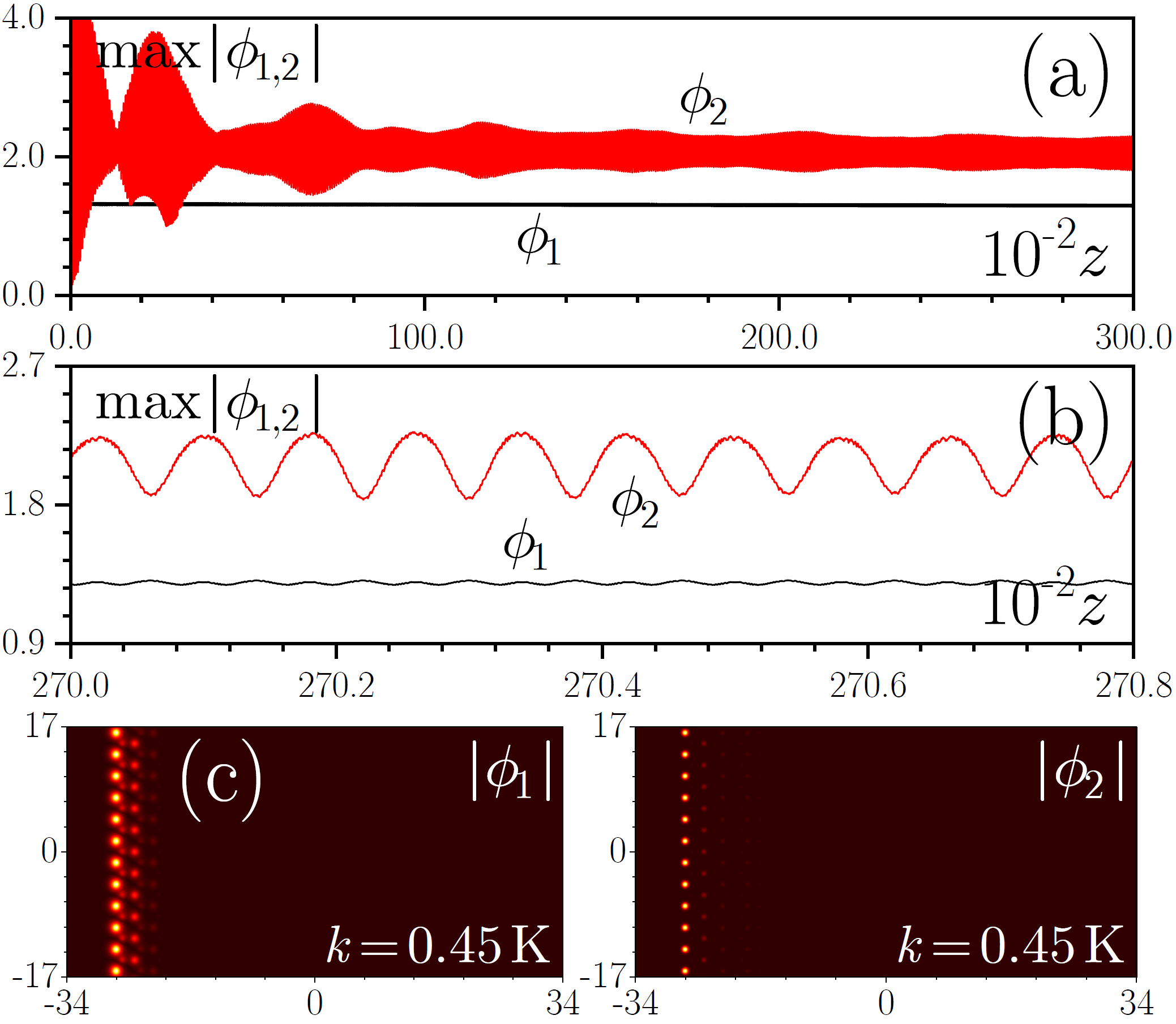}
	\caption{(a) Peak amplitudes of the carrier functions $\phi_1$ (black) and $\phi_2$ (red) versus distance $z$ and (b) zoom of this dependence at sufficiently large distances, when evolution of $\phi_2$ becomes periodic. (c) Profiles of the carrier functions $\phi_1$ and $\phi_2$. Here $\beta=7.6$, $k=0.45\textrm{K}$, while all other parameters correspond to Fig.~\ref{figure1}.}
	\label{figure2}
\end{figure}

The existence of small-amplitude quadratic bright solitons is possible for opposite signs of the effective dispersion $b^{\prime\prime}$ of chosen carrier edge state of the FF [see Fig.~\ref{figure1}(d) where it is shown that $b^{\prime\prime}$ can change its sign with $k$] and   of the effective nonlinear coefficient $\chi$. By multiplying Eq.~(\ref{exp-eq2}) by $\phi_2^*$ and subsequent integrating it over $\br$ and $z$, one can verify that $\chi$ is real. Its value is obtained from the expression (\ref{chi}) using the known linear Floquet edge state $\phi_1$ and  the respective solution of Eq. (\ref{exp-eq2}), $\phi_2$,  where $\phi_1^2$ enters as a "driving" field. The latter equation was solved numerically with zero initial conditions for $\phi_2$ function for representative experimentally realistic parameters of helical waveguide array indicated in the caption to Fig.~\ref{figure1}. Figures~\ref{figure2}(a) and (b) show typical evolution of peak amplitudes of the $\phi_1$ and $\phi_2$ functions at $k=0.45\textrm{K}$, corresponding to quasi-propagation constant $b=0.311=0.396\omega$. One can see that while the function $\phi_1$ always evolves in a periodic fashion due to its Floquet nature, the $\phi_2$ function approaches "steady" state, where it evolves practically periodically [see zoom in Fig. \ref{figure2}(b)], only after sufficiently long propagation distance. We use averaged amplitude $a_\textrm{av}=\langle |\phi_2|_\textrm{max}\rangle_Z$ as an indicator of the established SH field: almost unchanged $a_\textrm{av}$ indicates that the solution of Eq.~(\ref{exp-eq2}) converges to a $Z$-periodic function. In the case of Fig.~\ref{figure2}, where $\beta = 7.6$, the averaged amplitude gradually approaches the $a_\textrm{av}=2.07$ value. Corresponding representative profiles (on twelve $y$-periods) of the FF and SH carrier functions $\phi_1$ and $\phi_2$ at $z=30000$ are depicted in Fig. \ref{figure2}(c). One can see that the $\phi_2$ function is much more localized within waveguides, but at the same time its amplitude exhibits more pronounced $z$-oscillations. Notice that radiative losses for our parameters are nearly negligible (i.e. the amplitude does not notably decrease upon propagation).

The most important peculiarity of edge solitons in $\chi^{(2)}$ Floquet insulator, clearly distinguishing them from usual known $\chi^{(2)}$ solitons in materials with $z$-independent refractive index, stems from periodic $z$-dependence of the Floquet-Bloch mode $\phi_1$. Indeed, the right-hand side of Eq.~(\ref{mainSH}) can be viewed as a periodic driving force that can resonate with the eigenmodes of the SH wave. Such resonance is expected to occur when the frequency of the $\psi_1^2$ term (in our case, doubled quasi-propagation constant of the FF edge state $2b_{\alpha k}^{(1)}$), coincides for some band $\nu$ with quasi-propagation constant $b_{\nu k}^{(2)}$ of the eigenmode from the Floquet band of the SH wave. Taking into account that these constants are defined modulo $\omega$,  we thus gets the resonant condition
\begin{align}
\label{resnance}
b_{\nu k}^{(2)}=2b_{\alpha k}^{(1)}\quad (\textrm{mod}\,\, \omega)
\end{align}
We refer to this phase-matching mechanism as {\em Floquet phase matching}. Unlike in the case of a $z-$independent refractive index, where the phase mismatch $b_{\nu k}^{(2)}-2b_{\alpha k}^{(1)}$ is proportional to $\beta$ and is zero for $\beta=0$, now  the dependence on $\beta$ is richer, because the phase mismatch $\beta$ directly enters the equation (\ref{mainSH}) leading to a shift of the allowed bands for SH wave. This is shown in Fig. \ref{figure1}(c), where the resonance occurs for $\beta=\beta_{\rm res}$ where $\beta_{\rm res}\approx 6.71$ while $\beta=0$ corresponds to a nonresonant case. Under the condition of Floquet phase matching, the FF edge state resonantly excites the SH wave. For example, for the array depth of $p=11$ and for topological FF edge state with $b_{\alpha k}^{(1)}\approx0.311\approx0.396\omega$ at $k=0.45K$ such resonance is expected to occur with SH wave state at $b_{\nu k}^{(2)}\approx -0.163 \approx -0.208\omega$ when $\beta_{\rm res}\approx 6.71$, as illustrated in Fig~\ref{figure1}(c) (see lower group of bands). This is in full agreement with condition (\ref{resnance}). Notice that $\beta_{\rm res}$ depends on the Bloch momentum $k$.

\begin{figure}
	\centering
	\includegraphics[width=\linewidth]{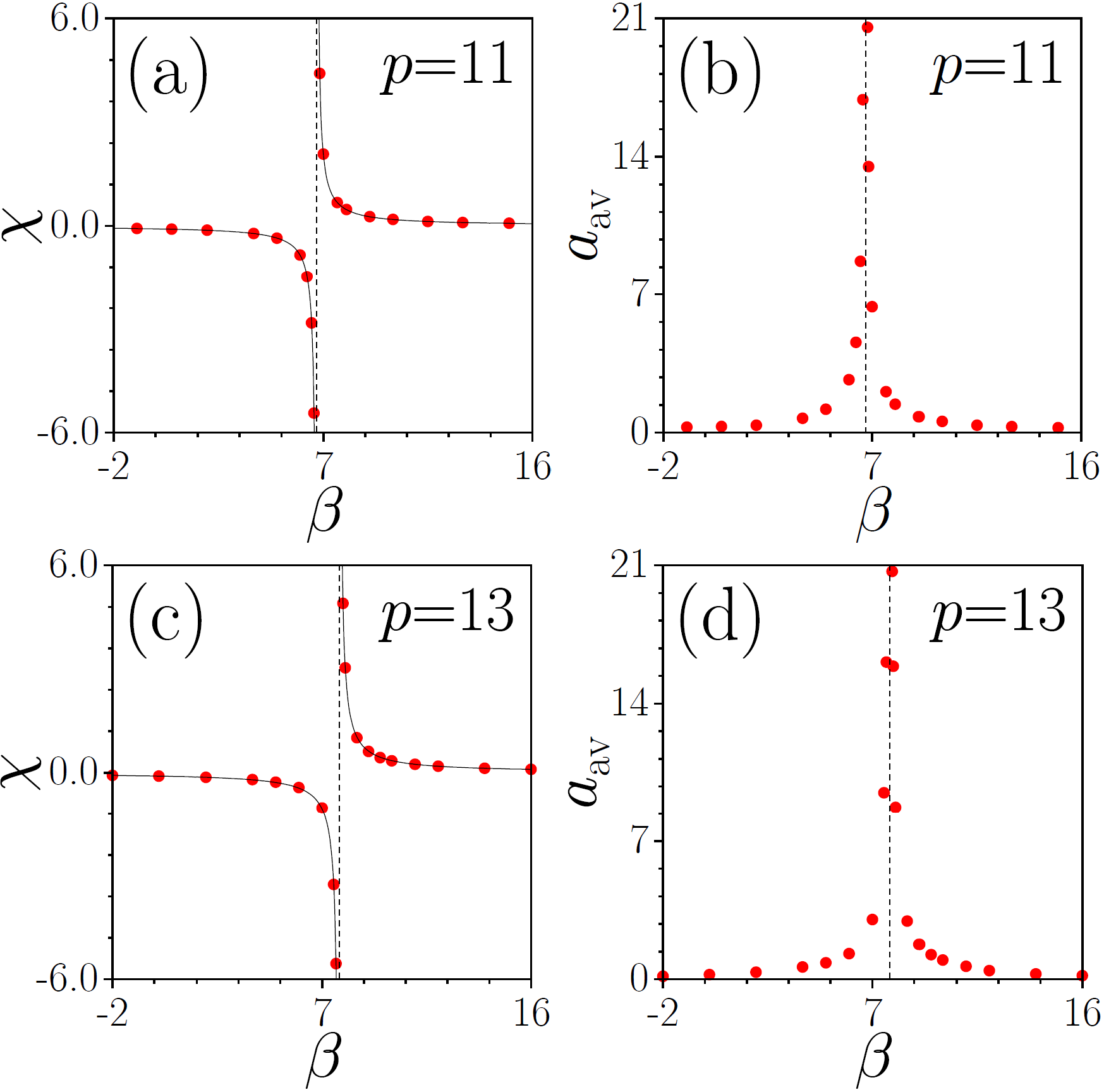}
	\caption{Approximate effective cubic nonlinear coefficient (a),(c) and $z$-averaged peak amplitude $a_{av}$ of the $\phi_2$ function (b),(d) versus phase mismatch $\beta$ at $k=0.45\textrm{K}$ and various array depths $p$. Vertical dashed lines indicate resonant phase mismatch values $\beta_\textrm{res}=6.71$ for $p=11$ (a),(b) and $\beta_\textrm{res}=7.74$ for $p=13$ (c),(d). Solid lines show approximations $\chi \approx 0.6032/(\beta-\beta_\textrm{res})$ in (a) and $\chi \approx 0.7845/(\beta-\beta_\textrm{res})$ in (c). The dependencies for $k=0.55\textrm{K}$ are practically identical, except for a slight shift in $\beta$.}
	\label{figure3}
\end{figure}

At, and near, Floquet phase-matching between the FF and SH waves, the multiple-scale model is not valid anymore. Phase matching is manifested in diverging effective cubic nonlinear coefficient, as shown by the red dots in Figs.~\ref{figure3}(a) and (c) for two different array depths $p=11$ and $p=13$ and for $k=0.45\textrm{K}$. The dependencies $\chi(\beta)$ calculated for other Bloch momentum $k=0.55\textrm{K}$ are very similar to those shown in Fig.~\ref{figure3}(a) and (c), except for a small shift in resonant value of $\beta$. The width of the resonance is determined by the width of the Floquet band of the SH wave spectrum and is very narrow due to flatness of the respective band (see Fig.~\ref{figure1}(c). The sign of the effective cubic nonlinear coefficient changes across the resonance: one has $\chi<0$ when $\beta<\beta_{\rm res}$, and $\chi>0$ when $\beta>\beta_{\rm res}$ [the dependence $\chi(\beta)$ can be accurately extrapolated by the law $\chi\propto 1/(\beta-\beta_\textrm{res})$, as shown by the solid lines and described in the caption of  Fig.\ref{figure3}]. This means that for a given sign of the dispersion coefficient (determined by the Bloch momentum $k$) bright solitons can be excited only in one of the two wide gaps in the Floquet-Bloch spectrum of the SH wave [Fig.~\ref{figure1}(c)]. Thus, for $k=0.45\textrm{K}$ the dispersion coefficient for the FF edge state $b''<0$ and therefore bright solitons can form only at $\beta>\beta_\textrm{res}$, where $\chi>0$. In contrast, for $k=0.55\textrm{K}$ the dispersion coefficient for the FF edge state $b''>0$ that allows soliton formation at $\beta<\beta_\textrm{res}$, where $\chi<0$. Thus, by properly selecting Bloch momentum $k$ and carrier edge state one can excite bright solitons for different values and signs of the material phase mismatch $\beta$ that adds considerable flexibility to the potential experimental realization of this system. The relation between the domains of existence of quadratic solitons and the bandgap structure resembles the well-known fact from the theory of gap solitons in straight waveguide arrays: due to different signs of the effective mass (effective diffraction) at the different edges of allowed band, (bright) soliton families can bifurcate only from one of the two edges, depending on the sign of the nonlinearity. Now, however, bifurcation picture is more sophisticated because effective diffraction coefficient is determined by the in-gap FF edge state, while the sign of effective nonlinearity depends on the interaction of the FF and SH waves in $z$-varying refractive index landscape. Since quasi-propagation constants of the bulk and edge states of the FF and SH waves depend on the array depth, for different $p$ values the Floquet phase matching conditions (\ref{resnance}) are met at different resonant values of phase mismatch $\beta_\textrm{res}$. It turns out that at fixed Bloch momentum the value $\beta_\textrm{res}$ increases with increase of $p$: one has $\beta_\textrm{res}\approx 6.71$ for $p=11$ and $\beta_{\rm res}\approx7.74$ for $p=13$ [compare Fig. \ref{figure3}(a) and (c)]. In the dynamics, the Floquet phase matching results in resonant growth of the averaged amplitude $a_\textrm{av}$ of the $\phi_2$ field, as shown in Figs.~\ref{figure3}(b) and (d).

\section{Edge solitons}

\begin{figure*}
	\centering
	\includegraphics[width=0.9\linewidth]{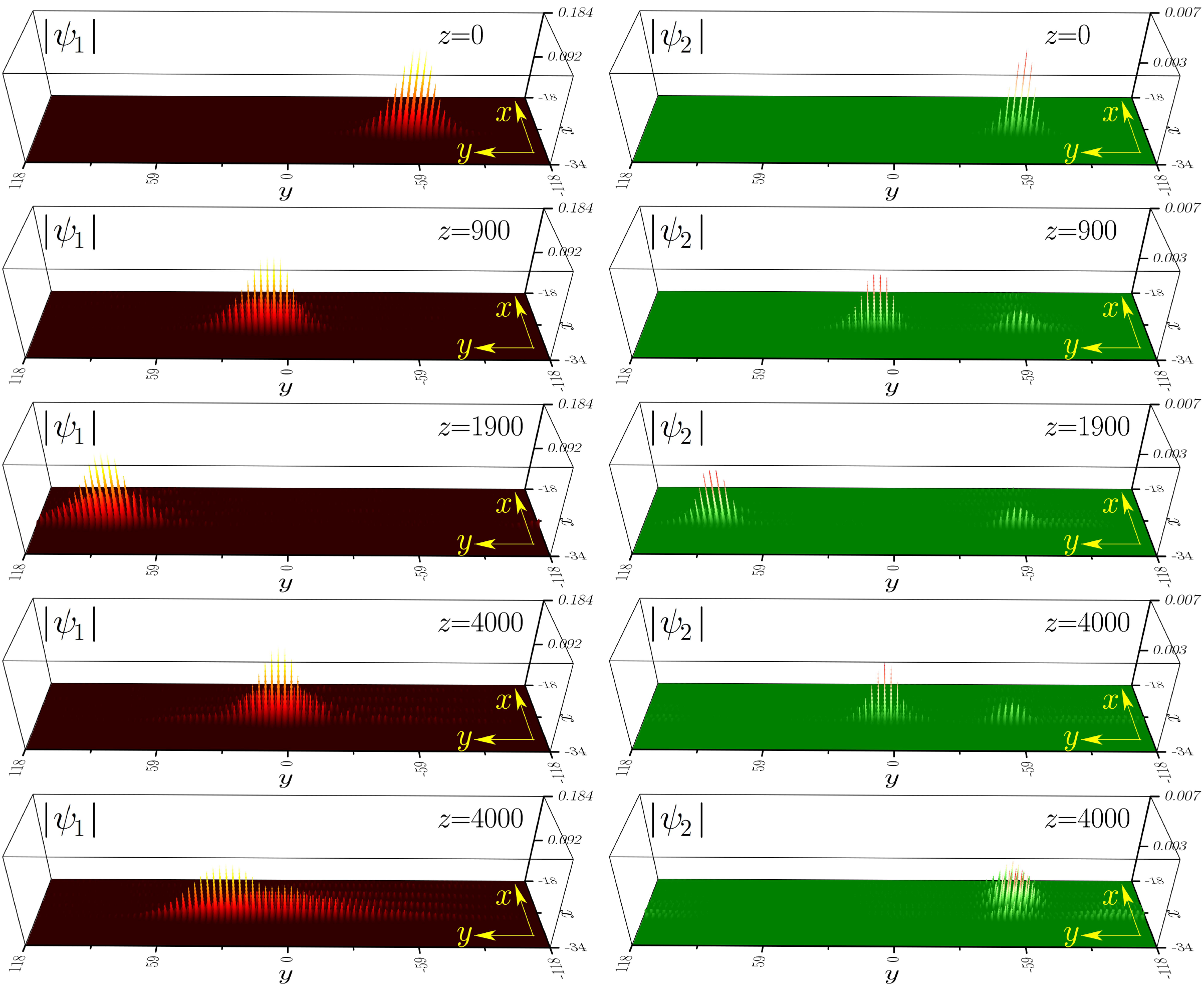}
	\caption{Propagation dynamics for $\beta=0$, $k=0.55K$, $\chi=-0.087$, $b''=+0.116$, $b_{nl}=-0.001$. Top four rows -- nonlinear propagation, bottom row -- linear propagation. Left column shows FF wave, right column shows SH wave. Notice different vertical scales in plots for FF and SH waves.}
	\label{figure4}
\end{figure*}

\begin{figure*}
	\centering
	\includegraphics[width=0.9\linewidth]{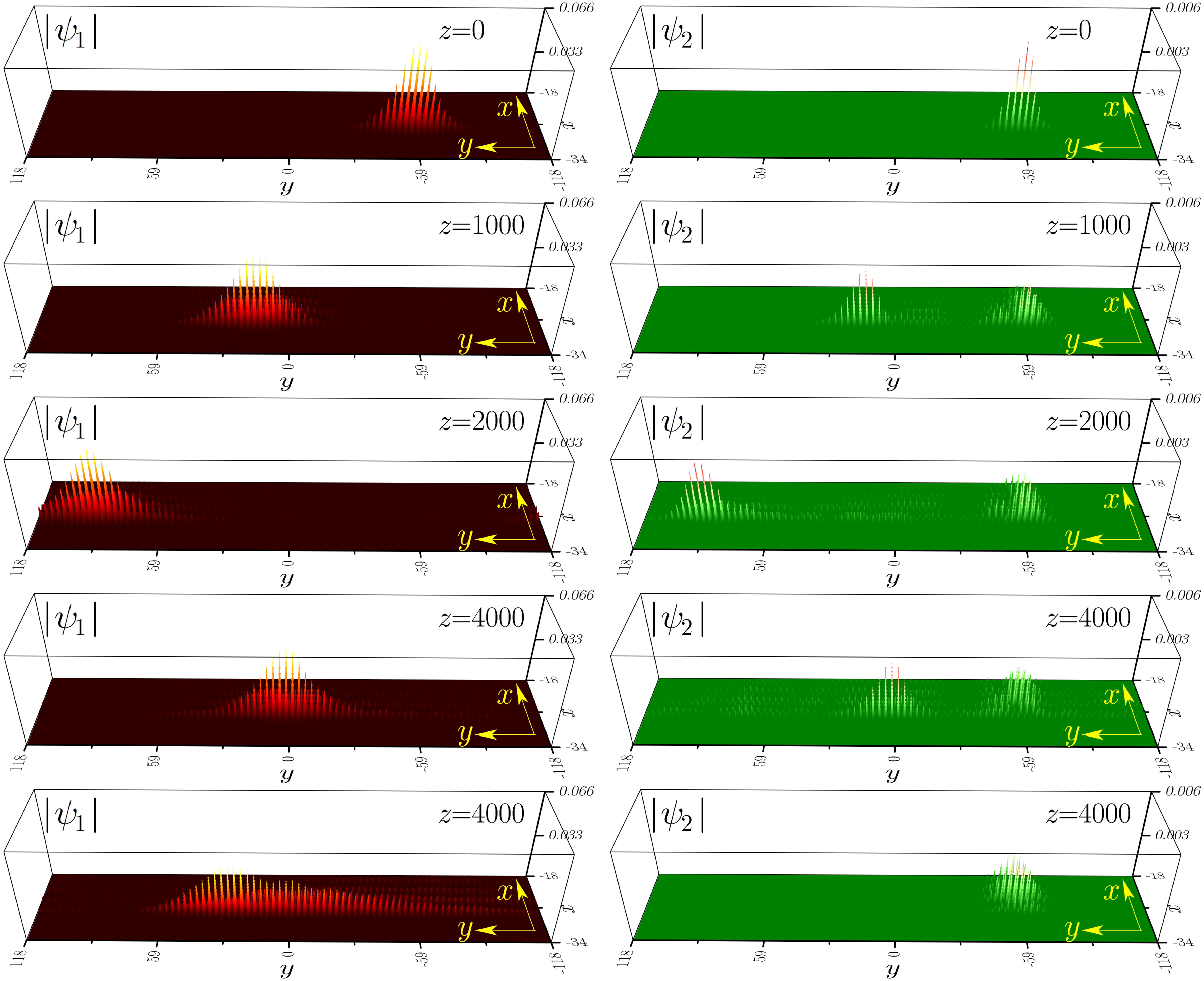}
	\caption{Propagation dynamics for $\beta=7.6$, $k=0.45K$, $\chi=+0.671$, $b''=-0.096$, $b_{nl}=+0.001$. Top four rows -- nonlinear propagation, bottom row -- linear propagation. Left column shows FF wave, right column shows SH wave. Notice different vertical scales in plots for FF and SH waves.}
	\label{figure5}
\end{figure*}

Equation~(\ref{NLS}) predicts that bright topological soliton propagating along the edge of the insulator in $\chi^{(2)}$ nonlinear medium is characterized by the envelope [see expansions (\ref{expan1}), (\ref{expan2}), assuming that corresponding FF and SH fields $\psi_1 \sim A$ and $\psi_2 \sim A^2$]: 
\begin{align} \label{SolEnv}
A = \left(\frac{2b_{nl}}{\chi}\right)^{1/2} \; \mathrm{sech}\left[\left(-\frac{2b_{nl}  }{b''  }\right)^{1/2}Y\right]e^{ib_{nl}z},
\end{align}
where $b_{nl}$ is a detuning of the quasi-propagation constant from its linear value $b$, arising due to nonlinearity. This envelope should provide accurate description as long as phase mismatch is not too close to resonant value $\beta_\textrm{res}$. One can see that in the case, when bifurcation of soliton occurs from linear FF edge state with $b''<0$, that is possible at $\beta>\beta_\textrm{res}$, where $\chi>0$, the nonlinear detuning of the quasi-propagation constant $b_{nl}$ should be positive . In contrast, for positive dispersion coefficient $b''>0$, at which soliton formation is possible at $\beta<\beta_\textrm{res}$, where $\chi<0$, the detuning $b_{nl}$ should be negative. This means that depending on Bloch momentum and phase mismatch solitons can bifurcate from the linear FF edge state in both directions within topological gap. Nonlinear detuning $b_{nl}$ should be small enough to ensure that total quasi-propagation constant remains in the topological gap and that envelope is broad enough and covers many $y$-periods.

To check the accuracy of the developed theory we propagate edge solitons, constructed as $\psi_1(\br,0)=A(y,0)\phi_1(\br,0)$ and $\psi_2(\br,0)=A^2(y,0)\phi_2(\br,0)$ in helical waveguide array, where nonlinear light evolution is governed by continuous Eqs.~(\ref{mainFF}) and (\ref{mainSH}). In Fig.~\ref{figure4}, rows 1-4, we illustrate the evolution of the edge soliton bifurcating from the red edge state branch [see Fig.~\ref{figure1}(b)] for the Bloch momentum $k = 0.55K$, corresponding to the dispersion coefficient $b''\approx +0.116$, and $\beta=0$, at which the effective nonlinear coefficient is negative $\chi\approx-0.087$ (this case corresponds to far off-resonant propagation). In this case, bright envelope (\ref{SolEnv}) corresponds to the negative quasi-propagation constant detuning, that we take here equal to $b_{nl}=-0.001$. In the left column of Fig.~\ref{figure4} we show evolution of the FF wave $|\psi_1|$, while right column demonstrates evolution of the SH wave $|\psi_2|$. One can see that both components are locked and move together along the edge without considerable modifications, even though the soliton traverses more than $100$ $y$-periods of the structure at $z=4000$ (notice that we work with huge, but finite $y$-window, so due to transverse displacement and periodic boundary conditions in $y$, after sufficiently long propagation distance the soliton may reappear from the other side of the window). Edge solitons are a hybrid asymmetric object: they are strongly localized across the interface, because they inherit the topological nature from the corresponding linear edge state, while along the interface they are localized due to nonlinearity. Soliton move with group velocity $v=-b^\prime$ dictated by the group velocity of corresponding edge state at selected Bloch momentum. Small reshaping on the SH component takes place only at the initial stage of propagation (this reshaping is due to the fact that our theory provides $z$-averaged approximation to the exact $z$-oscillating solution). It leaves small, practically immobile wavepacket in the SH wave at the initial launching position $y \approx -60$ that does not generate any appreciable FF wave and slowly diffracts into the bulk of the array (since there is no FF wave at this location anymore that was stimulating near-surface localization of the SH wave). As expected, most of the SH wave power remains locked to the FF wave and moves with it, forming stable edge soliton. Due to helicity of the waveguides, the amplitudes of FF and SH components undergo small $Z$-periodic oscillations upon propagation. Both fields $|\psi_1|$ and $|\psi_2|$ remain well-localized, illustrating a negligible influence of coupling to radiation into the bulk of the array due to topological origin of the mode on which the soliton is constructed. To prove that $\chi^{(2)}$ edge solitons are indeed supported by the nonlinearity, in the bottom row of Fig.~\ref{figure4} we show how the same input dramatically disperses in the linear medium. Without nonlinearity, we observe a strong asymmetric expansion of the FF wavepacket, whose peak amplitude substantially decreases. Since FF and SH waves are decoupled in the absence of nonlinearity, $\psi_2$ component remains immobile and slowly diffracts into the depth of the array.

To illustrate the existence of bright edge solitons for other sign of the dispersion coefficient, we consider also Bloch momentum $k = 0.45K$, corresponding to $b''\approx -0.096$, and take $\beta=7.6$, at which effective nonlinear coefficient $\chi\approx +0.671$ is positive. In this case, bright solitons correspond to positive nonlinear detuning $b_{nl}=+0.001$. Corresponding nonlinear propagation dynamics is illustrated in rows 1-4 of Fig.~\ref{figure5}. As in the previous case, both fields $|\psi_1|$ and $|\psi_2|$ remain well-localized upon propagation. Since in this particular case the phase mismatch $\beta = 7.6$ is taken closer to resonance at $\beta_{\rm res} \approx 6.71$ [see Fig. \ref{figure3}(a)], the reshaping of the SH wave is stronger and the fraction of SH power remaining in the launching point and slowly diffracting increases in comparison with Fig. \ref{figure4}. Still, the formation of the edge soliton steadily propagating along the interface over considerable distances is obvious (compare with linear propagation depicted in the last row of Fig. \ref{figure5}).

\begin{figure*}
	\centering
	\includegraphics[width=0.9\linewidth]{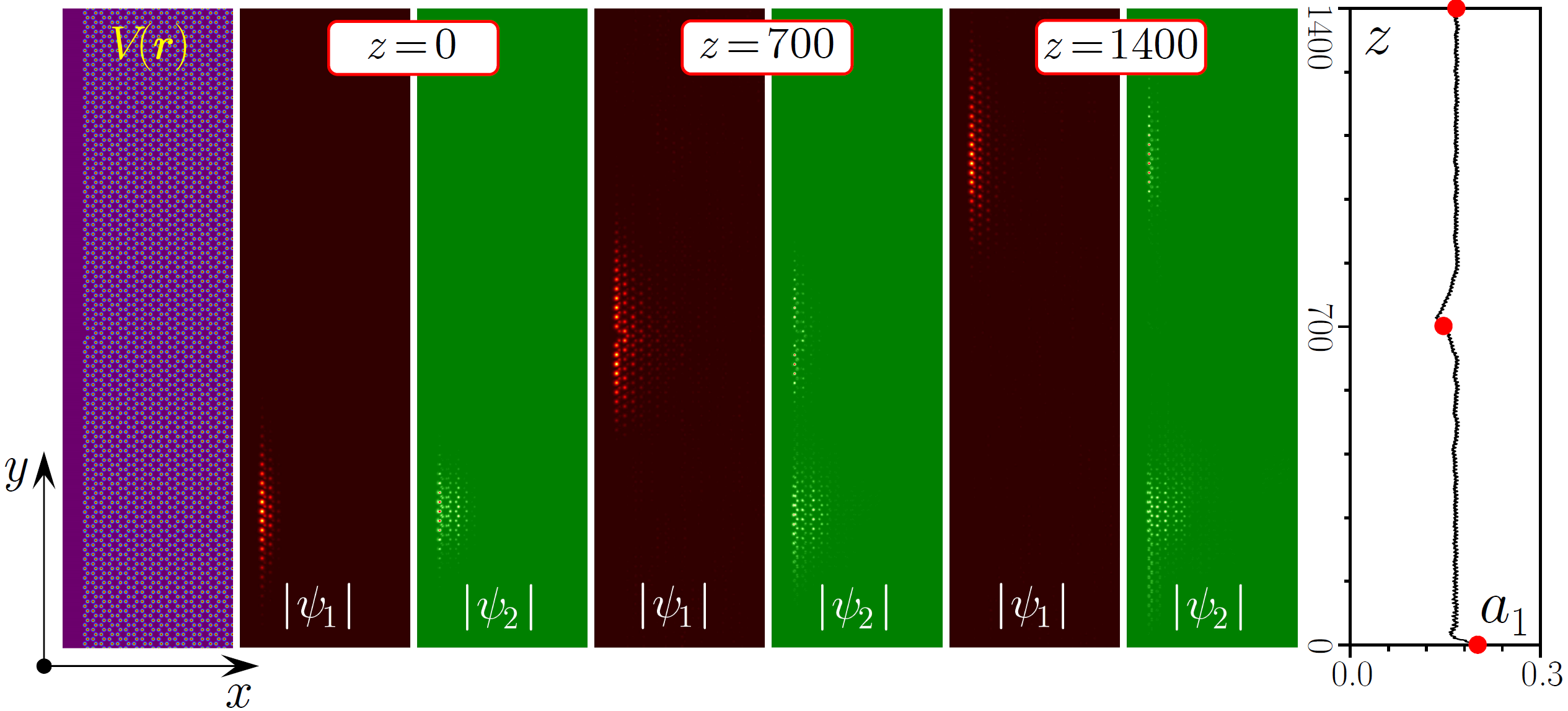}
	\caption{{Passage of parametric topological edge soliton from Fig. 4 through the defect in the form of missing waveguide. Profile of the array with a defect (left panel), field modulus distributions at different distances (central panels), and peak amplitude $a_1=\textrm{max}|\psi_1|$ of FF component versus $z$ (right panel). Shown distributions correspond to the red dots in the right panel.}}
	\label{figure6}
\end{figure*}

{To prove the topological origin of the multicolor edge solitons introduced here and to illustrate that our system with broken "time-reversal" symmetry provides topological protection manifested in the absence of backscattering on defects, we modeled the interaction of the edge soliton with strong defect in the form of missing waveguide at the zigzag edge of the array. As an input we used the same state at $\beta=0$, $k=0.55K$ as in Fig. \ref{figure4}. The wavepacket was launched at $y=-60$, sufficiently far from the defect at $y=0$ to allow it to reshape into exact traveling edge soliton at the moment of passage through the defect. The array with a defect and dynamics of soliton passage through it are illustrated in Fig. \ref{figure6}. One can see that soliton exhibits only local reshaping around the defect and that it restores its profile after passage of the defect. No appreciable backward reflection or radiation into the bulk is seen, neither in FF, nor in SH waves (the diffracting and non-moving fraction of SH wave around $y=-60$ is due to reshaping of the initial wavepacket into soliton and is not associated with backward reflection - it was discussed in Fig. \ref{figure4}). The peak soliton amplitude $a_1=\textrm{max}|\psi_1|$ (right panel) only slightly decreases when a soliton passes the defect and returns to the previous level sufficiently far from it. A similar level of topological protection was observed also for other momentum and phase mismatch values, e.g. for soliton from Fig. \ref{figure5}.}

\begin{figure*}
	\centering
	\includegraphics[width=0.9\linewidth]{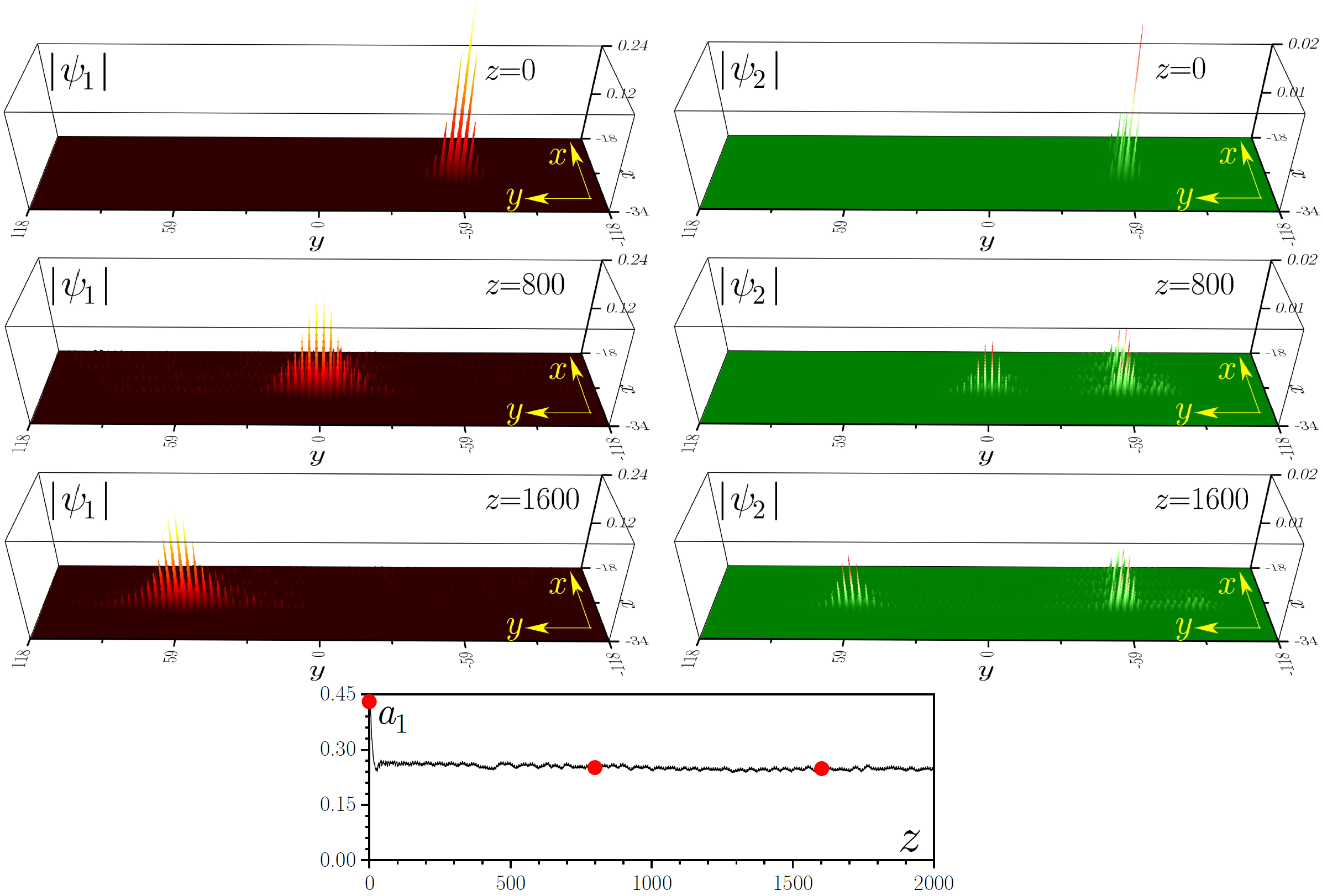}
	\caption{{Excitation of the edge soliton with narrow input at $\beta=0$, $k=0.55K$, $\chi=-0.087$, $b''=+0.116$, $b_{nl}=-0.005$. Left column shows FF wave, right column shows SH wave. Last row shows peak amplitude of FF wave versus $z$.}}
	\label{figure7}
\end{figure*}

{The theory of envelope solitons developed here implies that the envelope imposed on the edge state should be sufficiently wide to cover multiple $y$-periods of the array. However, we found that multicolor edge solitons can be efficiently excited even with relatively narrow input envelopes. This is illustrated in Fig. \ref{figure7}, where we used the same parameters as in Fig. \ref{figure4}, but increased the detuning of the quasi-propagation constant up to $b_{nl}=-0.005$. The resulting initial envelope is sufficiently narrow and covers only 4-5 $y$-periods. After the initial very fast stage of reshaping, where the wavepacket amplitude drops by approximately $30 \%$, the edge soliton forms that is slightly narrower than its counterpart from Fig. \ref{figure4} and that propagates with minimal amplitude oscillations, as shown in the last row of Fig. \ref{figure7}. The process of edge soliton excitation is, therefore, remarkably robust.}

\section*{Conclusions}

We reported the existence of topological Floquet solitons at the edge of honeycomb arrays of helical waveguides with $\chi^{(2)}$ nonlinearity. Such solitons have been described analytically and found numerically in a continuous model. We have shown that, away from so-called Floquet phase-matching resonance, the envelope of such solitons can be described by a single effective cubic nonlinear Schr\"odinger equation with a nonlinear coefficient dictated by the global phase mismatch existing between FF and SH waves, which includes a geometrically-induced shift. The Floquet phase-matching resonance occurs when the overall phase-mismatch vanishes, a condition that depends on the location of the allowed quasi-propagation constant bands of the SH and FF edge states. Floquet edge solitons obtained here are robust, they propagate along the edge over hundreds of the array periods remaining well-localized and showing no appreciable power losses into bulk modes.

\medskip
\textbf{Acknowledgements} \par %delete if not applicable))
Y.V.K., A.S. and S.K.I. acknowledge funding of this study by RFBR and DFG according to the research project no. 18-502-12080 and SZ 276/19-1. V.V.K. acknowledges financial support from the Portuguese Foundation for Science and Technology (FCT) under Contract no. UIDB/00618/2020. Y.V.K. and L.T. acknowledge  support from the Government of Spain (Severo Ochoa CEX2019-000910-S), Fundaci\'o Cellex, Fundaci\'o Mir-Puig, and Generalitat de Catalunya (CERCA program).

\medskip
\textbf{Disclosures} \par %delete if not applicable))
The authors declare no conflicts of interest.

% References

\end{document}